# A NEW MAIN INJECTOR RADIO FREQUENCY SYSTEM FOR 2.3 MW PROJECT X OPERATIONS *

J. Dey#, I. Kourbanis, Fermilab, Batavia, Illinois 60510, U.S.A.

*Abstract*

For Project X [1] Fermilab Main Injector will be required to provide up to 2.3 MW to a neutrino production target at energies between 60 and 120 GeV. To accomplish the above power levels 3 times the current beam intensity will need to be accelerated. In addition the injection energy of Main Injector will need to be as low as 6 GeV. The current 30 year old Main Injector radio frequency system will not be able to provide the required power and a new system will be required. The specifications of the new system will be described.

## SYSTEM SPECIFICATIONS

The high level specifications of the new Main Injector radio frequency system are outlined in Table 1. The RF frequency will remain the same in order to minimize changes in instrumentation. The frequency range has increased to 490 kHz in order to accommodate injection at 6 GeV. The maximum acceleration rate is 240 GeV/sec and given the revolution period of Main Injector, requires 2.7 MV for acceleration. The peak RF power is 6.2 MW and the average 3.0 MW since the RF is off for half the Main Injector Cycle. The maximum required voltage is determined by the bucket area allowing enough overhead to operate with one/two cavities down. The average beam current is 2.3A while the fundamental RF current can be as high as 4.1A after transition.

## RF CAVITY DESIGN

The chosen R/Q of the new cavities was a compromise between beam stability and power consumption. Here we assumed 240KV/cavity for a total of 20 cavities and a Q of 10000.

Table 1: RF System Specifications

| Parameter | Value | Units |
|---|---|---|
| Frequency | 52.617-53.104 | MHz |
| Max. Acc. Rate | 240 | GeV/sec |
| Frequency Slew Rate | 1.6 | MHz/sec |
| Acceleration Voltage | 2.7 | MV |
| Peak Beam Power | 6.2 | MW |
| Average Beam Power | 3 | MW |
| Peak Voltage | 4.7 | MV |
| Average Beam Current | 2.3 | A |
| Fundamental RF Current | 3.7-4.1 | A |

The Robinson Stability [2] factor for different levels of fundamental RF feedback [3] vs. R/Q is shown in the top plot in Fig. 1. The bottom plot in the same figure shows the power per station vs. R/Q.

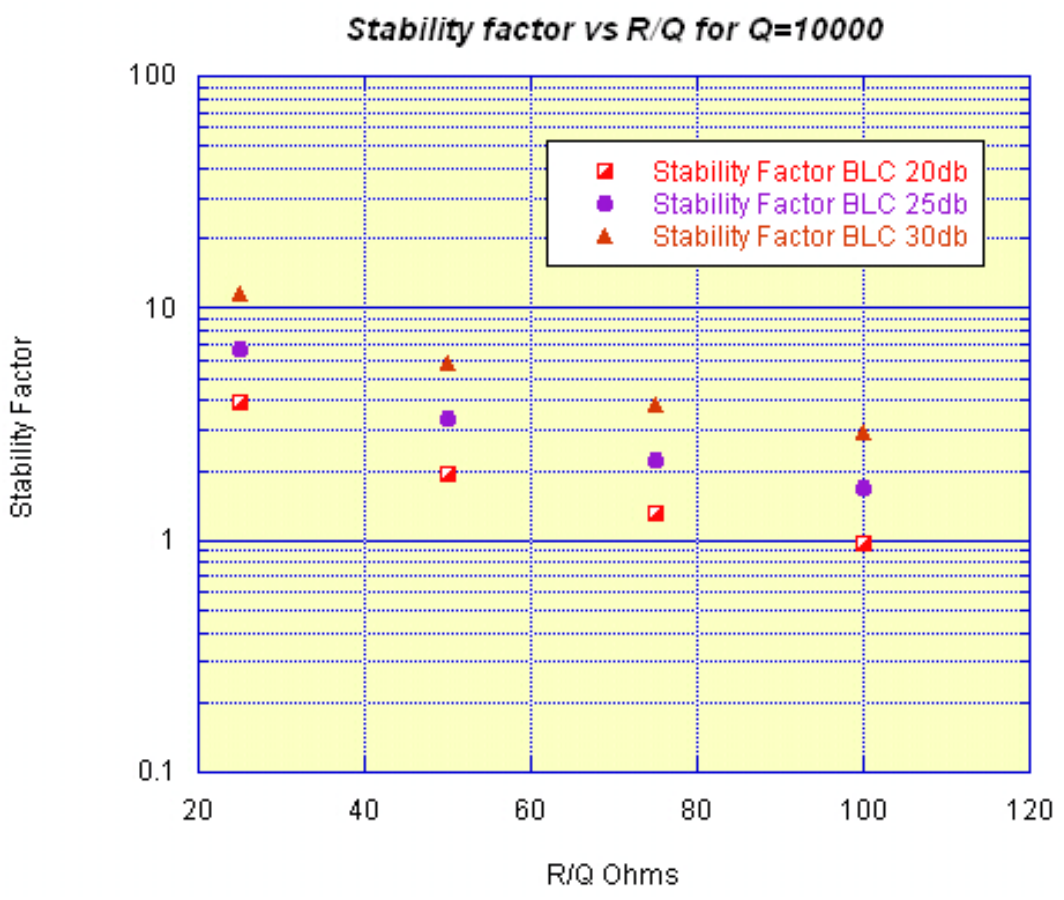

#dey@fnal.gov

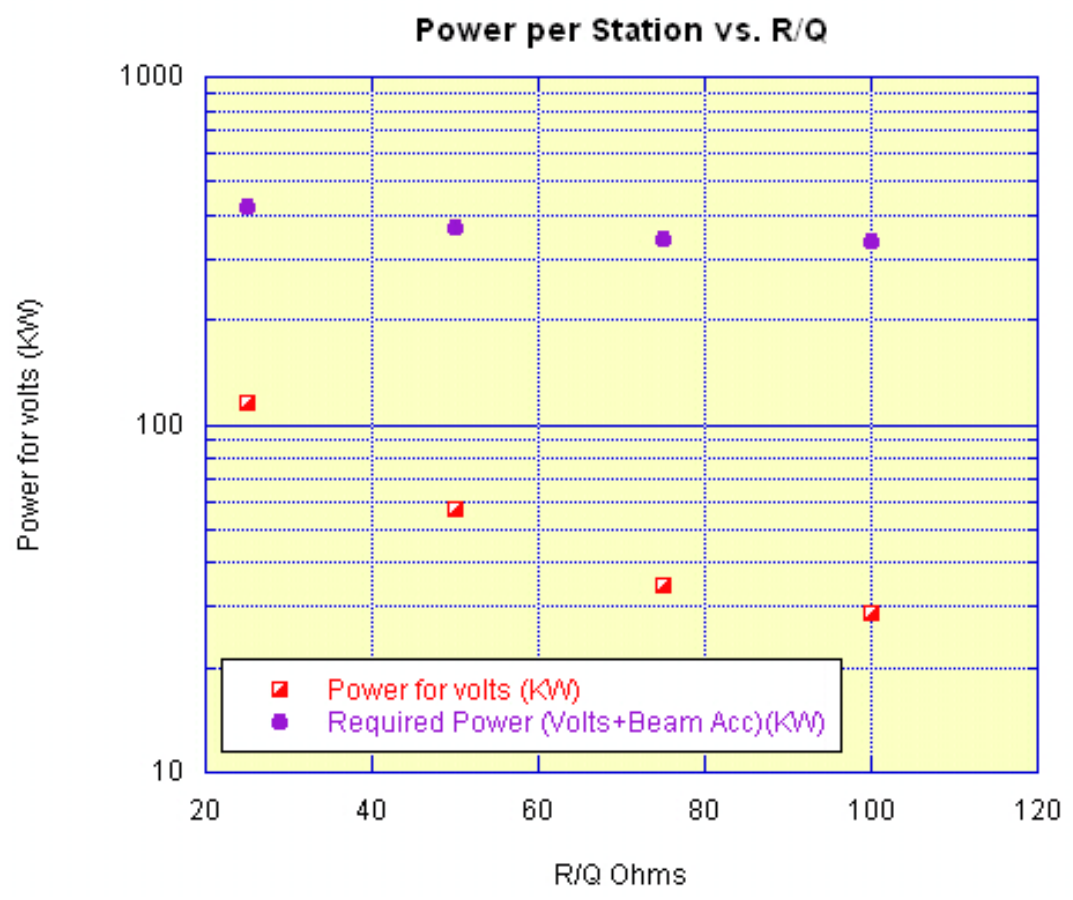

Figure 1: Robinson Stability factor vs. R/Q (top) and power per RF station vs. R/Q (bottom).

From the plots in Fig. 1 we can see that an R/Q around 50 Ohms is a good compromise. We can maintain a stability margin greater than 2 and dissipate less than 50 KW per cavity.

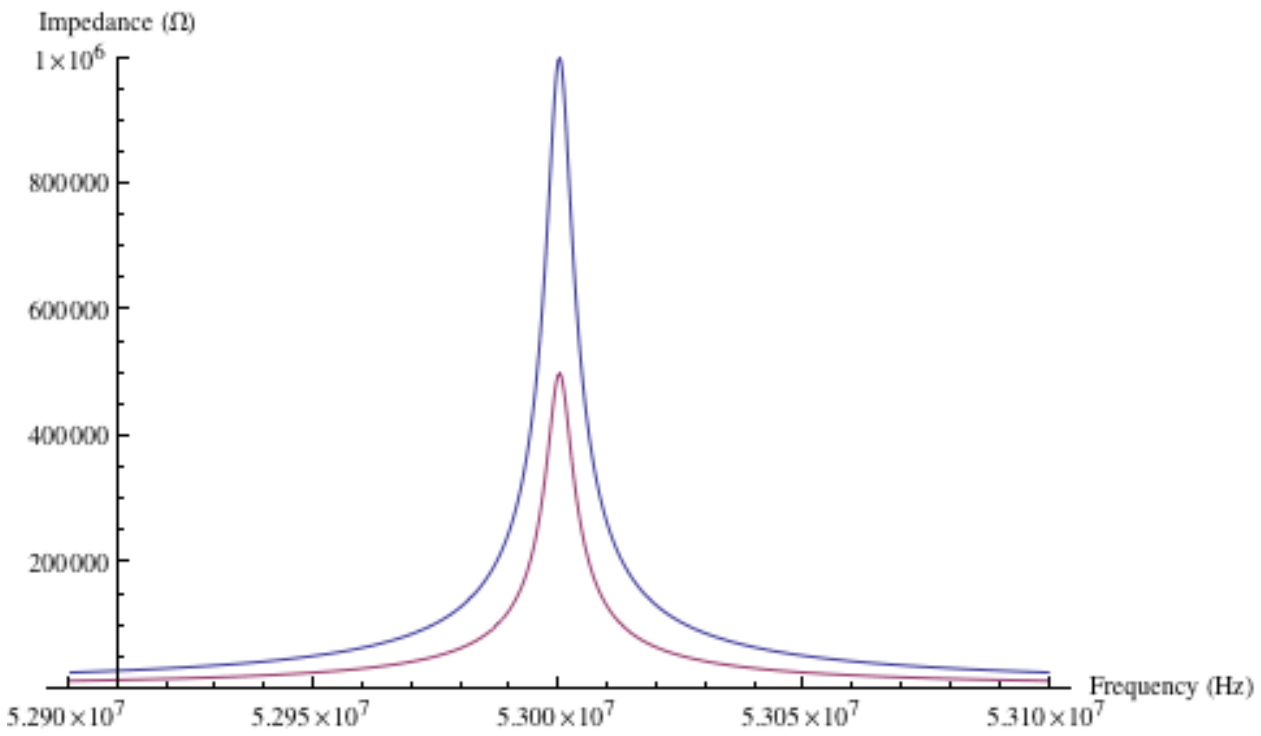

Figure 2: Blue Trace: R/Q = 100 Ω and Q = 10,000, Red Trace: R/Q = 50 Ω and Q = 10,000.

### *Cavity R/Q and Transients*

In Figure 2, Impedance vs. Frequency is plotted for a cavity resonating at 53 MHz and a Q of 10,000. The subtle difference is that the red trace has an R/Q that is half that of the blue trace. This becomes of great importance because the Main Injector has a revolution frequency of approximately 90 kHz. The y-axis in Figure 2 is placed at the first lower revolution line and illustrates that the impedance for the lower R/Q cavity is half that of the impedance of the higher R/Q cavity. Thus, the R/Q of a cavity plays an important role in lowering the transient beam loading for a given cavity.

## CAVITIES

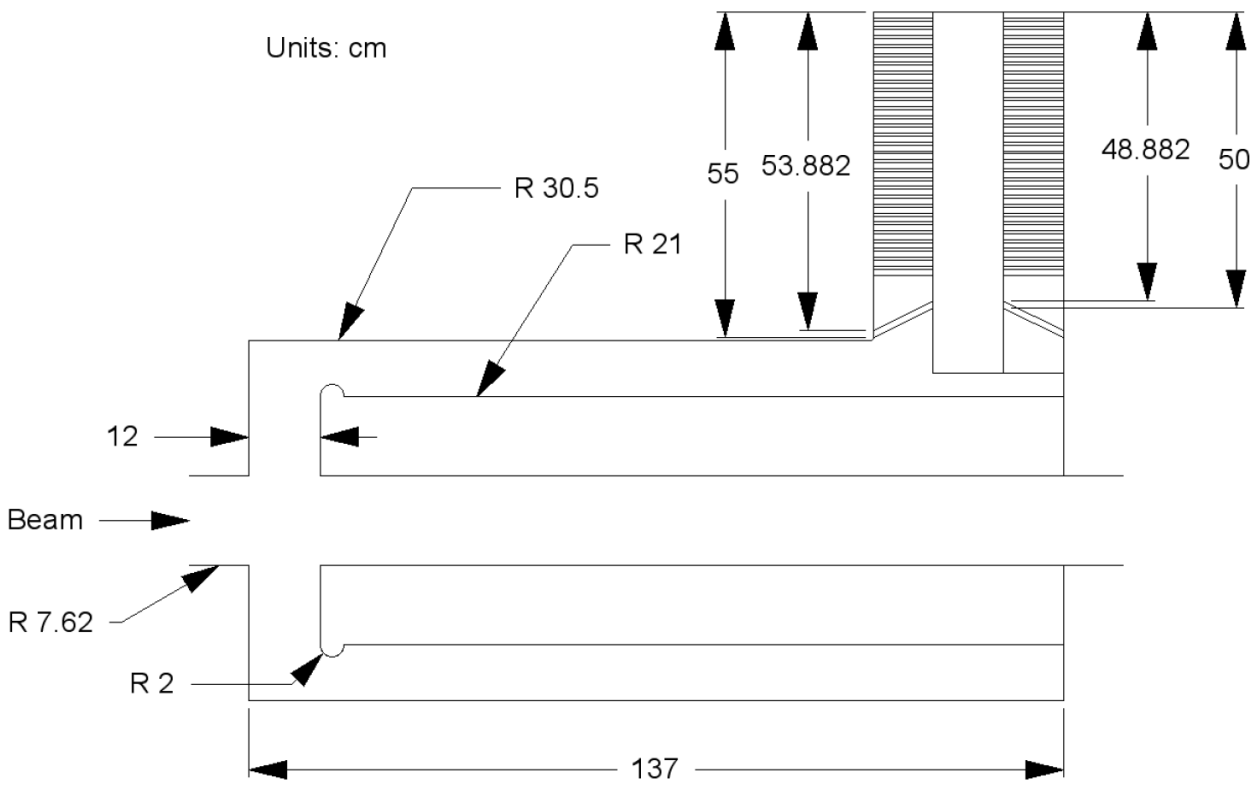

Figure 3: Project X Main Injector 53 MHz Cavity I

The initial criteria for the new 53 MHz Project X cavity are that it should be a quarter wave coaxial resonator with a single accelerating gap, have an R/Q of roughly 50 Ω, be made of OFHC copper, and have a perpendicularly biased ferrite tuner. The diameter of the Main Injector beam pipe is 15.24 cm. These specifications resulted in the Project X Main Injector Cavity I shown in Figure 3. For practical reasons, Cavity I transformed into the cavity shown in Figure 4. The advantages of Cavity II are that it requires only a single vacuum ceramic (tuner and driver will be at atmospheric pressure), the conical shape gives support to the inner conductor lever arm, and lastly, the cavity has a larger tuning range. Detailed electromagnetic simulations [4] of both cavity designs have been performed using finite element codes.

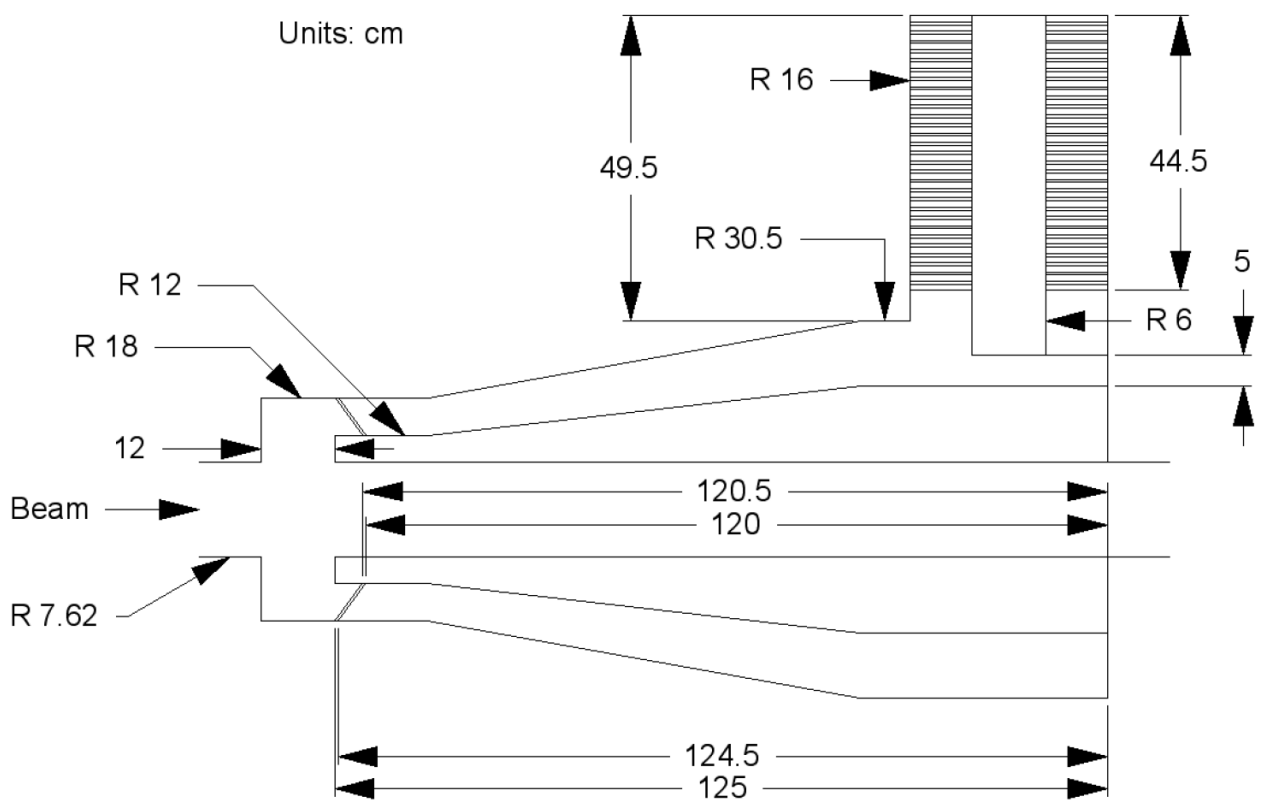

Figure 4: Project X Main Injector 53 MHz Cavity II

## CAVITY CALCULATIONS

$\phi$s = 33.6327 deg

Fundamental Beam Current (Ib) = 4.51042 A

R/Q = 50 Ω

Q = 10000

Cavity Voltage (Vc) = 240. kV

Cavity Power Loss per Cavity = $57.6 \times 10^3$ W

Total Apparent Power = $558.417 \times 10^3$ VA ∠ 50.209 degrees

Total Current (Ig) = 4.65347 A ∠ 50.209 degrees

Percent of Induced Mode Compensated = 26.4 dB = 95.2137 %

Robinson Stability factor = 4.01439

Figure 5: Calculated Parameters for a Cavity Operating at a Kinetic Energy of 30 GeV.

The Fundamental Beam Current in Figure 5 was found by doing a Fourier Analysis of the Main Injector filled with $1.6 \times 10^{14}$ protons. Operating at 30 GeV was chosen because this is at the point of when peak power is required. Figure 6 is the Vector Diagram of how one arrives at the Total Current of 4.65 A delivered to the gap of the cavity. A Robinson Stability factor of 4 was found and can be achieved by doing Direct RF Cavity Feedback with a Closed Loop Gain of 20.89.

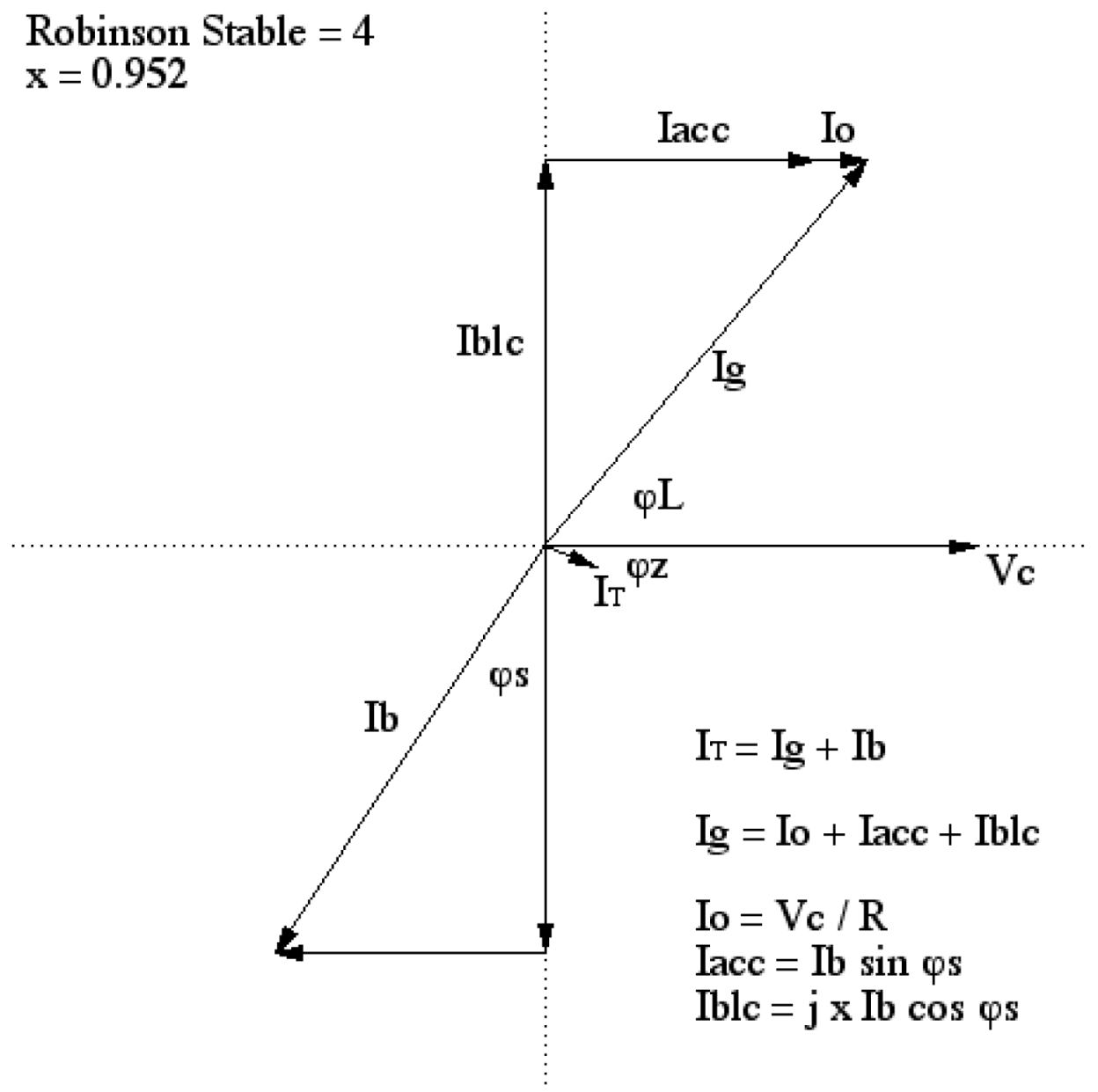

Figure 6: Vector Diagram of Current for a Cavity Operating at a Kinetic Energy of 30 GeV.

## POWER SOURCE

From Figure 5, the power source for the new cavity would be required to supply 558 kVA of total apparent power. Another criteria is that the power source needs to be able to operate at both 53 MHz and 106 MHz because we would also like to use the same tube for a second harmonic cavity system. The last criterion is that the new tube should have plenty of plate dissipation capabilities. After much consideration, it was decided that the tube of choice is the Eimac 8973 power tetrode. Some key qualities of the tube are that it has an output power capability greater than 1 MW, an operating frequency up to 110 MHz and a plate dissipation capability of 1,000 kW. Figure 7 shows the tube that was purchased from Communications and Power Industry (CPI) and received in September of 2010.

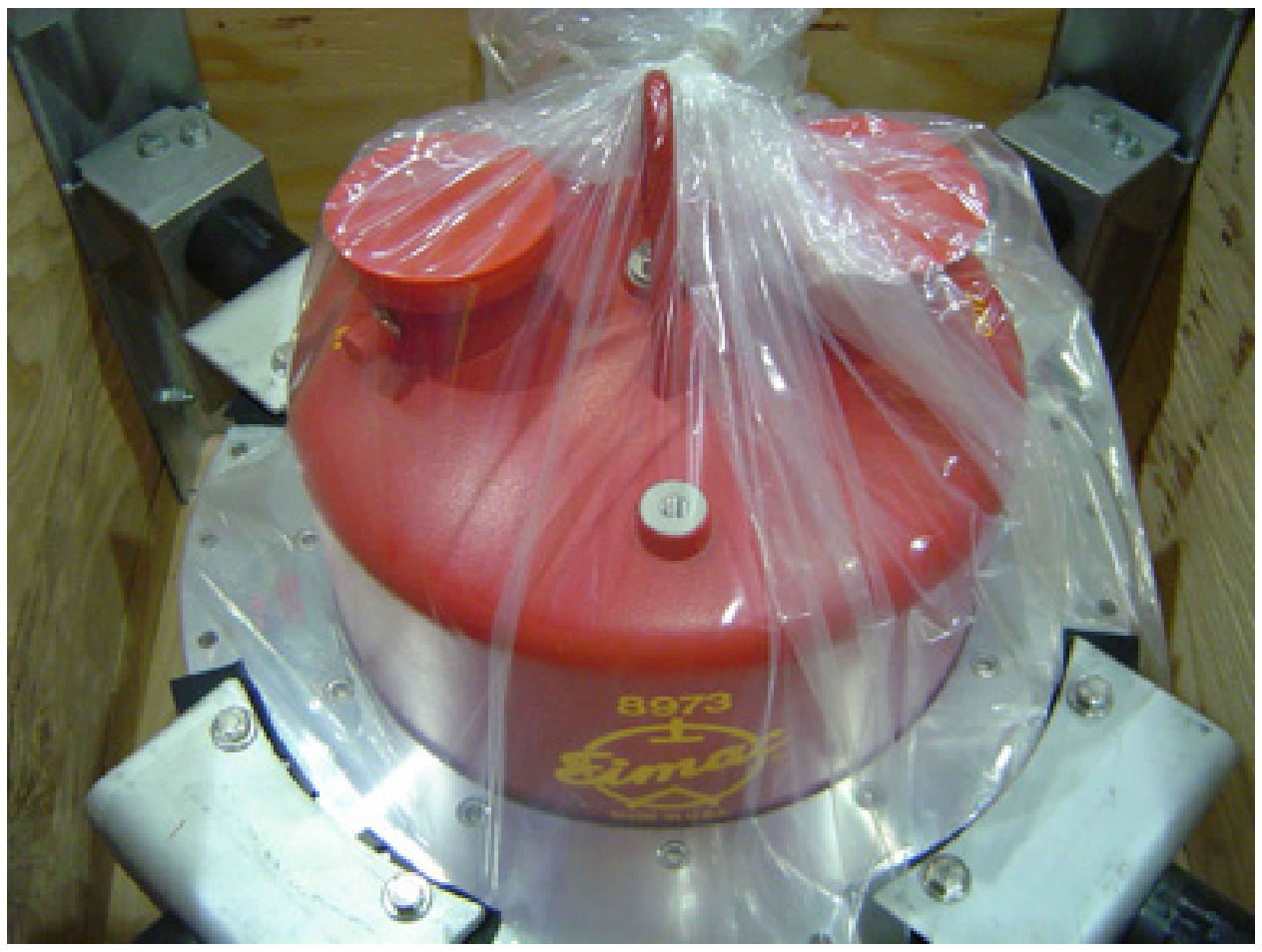

Figure 7: Eimac 8973 Power Tetrode.

### *Eimac 8973 Tube Calculations*

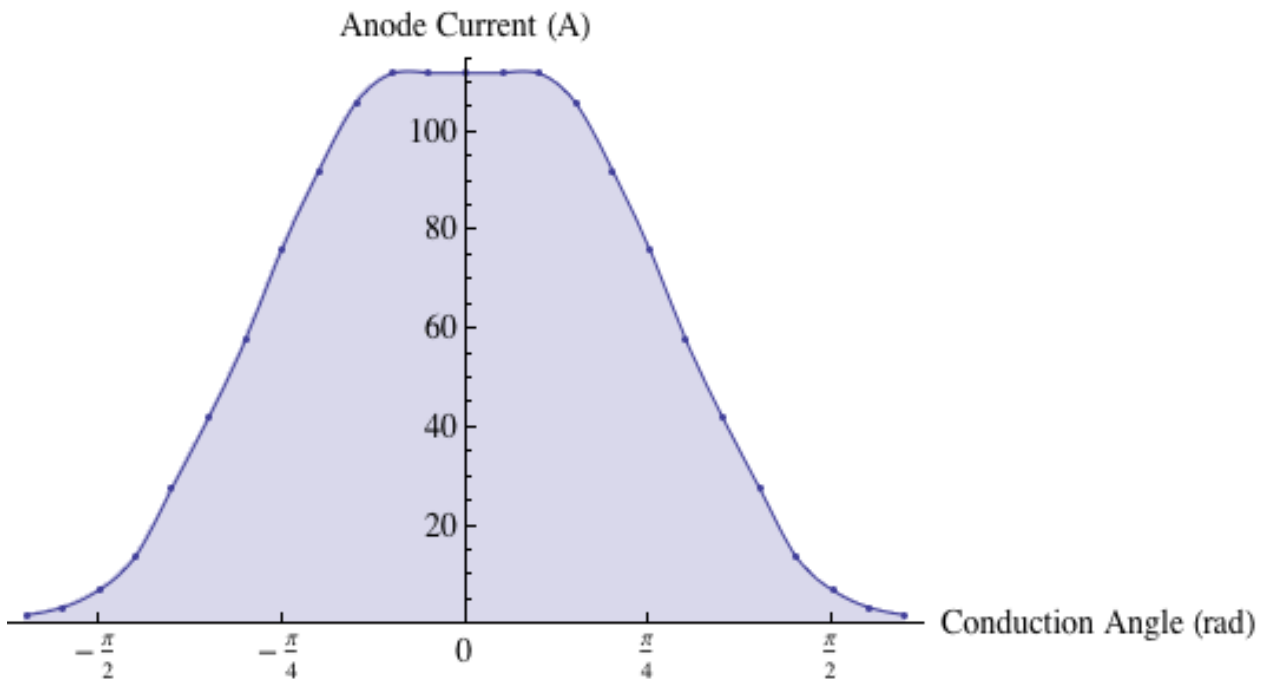

Figure 8: Eimac 8973 Anode Current vs. Conduction Angle.

In using the Eimac 8973 Tube curves for a grounded grid and screen to grid voltage of 1250 V, one can find the plot for Anode Current vs. Conduction Angle shown in Figure 8. The control grid is DC biased to -250 V and cathode driven with a peak cathode RF voltage ($V_{cath}$) of 240 V. The DC anode voltage ($V_{DC}$) is 22 kV and the anode RF voltage ($V_{RF}$) swing is 20 kV. The step-up ratio for the cavity is assumed to be 12:1. In knowing that the RF frequency is 53 MHz, one can do Fourier Analysis of the plot in Figure 8 to find that the DC current ($I_{DC}$) is 35.36 A and the peak RF current ($I_{RF}$) is 55.94 A. For a cathode driven tube, the input RF current is equal to the output RF current. The Power Calculations are shown below.

$$P_{DC} = V_{DC} I_{DC} = (22kV)(35.36A) = 777.9kW$$

$$P_{RF} = \frac{V_{RF} I_{RF}}{2} = \frac{(20kV)(55.94A)}{2} = 559.4kW$$

$$P_{cath} = \frac{V_{cath} I_{RF}}{2} = \frac{(240V)(55.94A)}{2} = 6.7kW$$

$$Efficiency = \frac{P_{RF}}{P_{DC}} 100 = 71.9\%$$

$$Plate\ Dissipation = P_{DC} - P_{RF} = 218.5kW$$

The RF Power of 559.4 kW meets the Total Apparent Power requirement shown in Figure 5 for 30 GeV operations. Also, the calculated Plate Dissipation of 218.5 kW is well below the tube rating of 1,000 kW.